\documentclass[12pt]{article}

\usepackage{amsfonts,amssymb}

\input epsf

\def\Im{\mbox{Im}}
\def\Re{\mbox{Re}}

\title{$W$-pair production in modified perturbation theory}
\author{M. L. Nekrasov \\
\small{\it Institute for High Energy Physics, 142281 Protvino, Russia}}

\date{}
\begin{document}

\maketitle

\begin{abstract}
We examine capabilities of the modified perturbation theory (MPT)
for description of the processes with productions and decays of
fundamental unstable particles. We calculate total cross-section for
$e^{+} e^{-} \to \gamma,Z \to W^{+} W^{-} \to 4f$ in a model with the
Dyson-resummed and with the MPT-expanded up to the NNLO Breit-Wigner
factors, and compare the outcomes. At the ILC energies a coincidence of
the outcomes is detected with precision better than one per-mille.

%\keywords{unstable particles; modified perturbation theory.}
\end{abstract}

%\ccode{11.15.Bt, 13.66.Jn} % 13.38.-b, 14.65.Ha

\section{Introduction}\label{int}

An investigation of the $W$-pair production plays an important role
in testing the Standard Model and searching for physics beyond. At
the International $e^{+} e^{-}$ Linear Collider (ILC) \cite{ILC}
the accuracy of the appropriate measurements is planned at the
per-mille level. This implies that the theoretic calculations of
the cross-section for $W$-pair production must be carried out with
the per-mille precision or better. To achieve this, the
calculations in the next-to-next-to-leading order (NNLO) are
required and, in addition, the resonant contributions of unstable
particles must be treated with the appropriate accuracy without
violation of the gauge cancellations.

A simultaneous implementation of the above requirements is a very
difficult task, unresolved so far. Really, the approach of the double
pole approximation (DPA) successfully applied at LEP2, provides only the
NLO precision and in the resonant region only \cite{LEP2}. So the DPA is
unable to solve the problem \cite{CMS1}. The pinch-technique method and
the background-field formalism, another for a long time developed
approaches \cite{pinch,BFM}, seemingly can provide the NNLO precision.
But for the maintenance of the gauge cancellations, they require a lot
of calculations of extra contributions that appear formally beyond the
necessary precision, which is impractical \cite{Ditt}. So, at present
the main hopes are pinned on the approach of ``complex-mass scheme''
(CMS) which avoids the mentioned difficulties \cite{CMS1,CMS2}.
Nevertheless, the CMS due to intrinsic problem related to unitarity,
provides the NLO precision only \cite{CMS2}. So, alternative approaches
are required.

A modified perturbation theory (MPT) \cite{Tkachov,EPJC,MPT} is a
probable candidate to become such an approach. Its main feature is
the direct expansion of the probability instead of amplitude in
powers of the coupling constant with the aid of
distribution-theory methods. The latter methods allow one to
impart a well-defined meaning to the resonant contributions in the
expansion of probability. A condition of asymptoticity (and
completeness) of the expansion must ensure the gauge
cancellations. In the case of pair production of unstable
particles, the most-elaborated description of the method was given
in \cite{MPT}. In \cite{N-tt,Ntt}
the convergence properties of the MPT expansion were tested in a
model related to the top-quark pair production. At the ILC energies, a
good convergence of the MPT series was detected with the precision of
the description within the NNLO up to a few per-mille.

In this Letter we present results of similar investigation in a model
for $W$-pair production in $e^{+} e^{-}$ annihilation. We emphasize that
a separate analysis in this case anyway must be made since this process
is planned for the ILC, and since this process has peculiar features.
Namely, in the case of $W$-pair production there appear the $t$-channel
contributions, which in the cross-section result in the contributions
with a more complicated analytic structure (the additional logarithm,
see e.g.~\cite{Gentle}). Moreover, there appear the large gauge
cancellations arising even in the Born approximation from contributions
of longitudinally polarized $W$-bosons. In particular, the mentioned
cancellations reach one order of magnitude at 400 GeV and two orders at
1 TeV \cite{Denner}. So, one should check the behavior of the MPT in the
presence of such large cancellations.

In the next section, we detail the model in which we carry out
calculations. The numerical outcomes are presented in Sec.~\ref{num}. In
Sec.~\ref{Summary}, we discuss the results.

\section{The total cross-section and a model for $W$-pair
production}\label{not}

The total cross-section of $W$-pair production in $e^+ e^-$
annihilation has the form of a convolution of the hard-scattering
cross-section with the flux function \cite{Gentle,Denner},
\begin{equation}\label{not1}
\sigma (s) = \int_{s_{\mbox{\tiny min}}}^s \frac{\mbox{d} s'}{s}
\: \phi(s'/s;s) \> \hat\sigma(s')\,.
\end{equation}
(The angular distributions may be described in the MPT, as well, but we
do not discuss this option here.) The flux function $\phi$ stands for
contributions of nonregistered photons emitted in the initial state.
Below we consider $\phi$ in the leading-log approximation (see details
e.g.~in \cite{Gentle}).
The hard-scattering cross-section is given by an integral over the
virtualities of unstable particles,
\begin{equation}\label{not2}
\hat\sigma (s) = \!\!\!\!\!\!
 \int\limits_{\quad {\displaystyle\mbox{\scriptsize $s$}}_
                    { 1 \mbox{\tiny min} } \atop
              \quad {\displaystyle\mbox{\scriptsize $s$}}_
                    { 2 \mbox{\tiny min} }}
            ^{\!\!\infty}   \!\!\!\!\!\!\!\!
 \int\limits^{\infty} \mbox{d} s_1 \, \mbox{d} s_2 \;
 \theta(\!\sqrt{s}-\!\sqrt{s_1}-\!\sqrt{s_2}\,)
 \sqrt{\lambda (s,s_{1},s_{2})}\;\Phi(s;s_1,s_2)
 \left(1\!+\!\delta_{c}\right)
 \rho(s_{1}) \> \rho(s_{2})\,.
\end{equation}
Here $s_{1\,\mbox{\scriptsize min}}$ and $s_{2\,\mbox{\scriptsize
min}}$ are minima of the virtualities,
$\sqrt{s_{1\,\mbox{\scriptsize min}}} +
\sqrt{s_{2\,\mbox{\scriptsize min}}} = \sqrt{s_{\mbox{\scriptsize
min}}}$. The $\lambda = [s \!-\!(\sqrt{s_1} +\! \sqrt{s_2}\,)^2]
[s \!-\!(\sqrt{s_1} - \! \sqrt{s_2}\,)^2]$ is the kinematic
function. The $\theta$-function and the square root of $\lambda$
constitute a kinematic factor. The $\rho(s_i)$ are Breit-Wigner
(BW) factors of $W$-bosons. The $\Phi$ and the factor
$(1\!+\!\delta_{c})$ stand for the amplitude squared and soft
photon contributions. Since $\Phi$ corresponds to one-particle
irreducible contributions, it has no singularities on the
mass-shell of $W$-bosons. On the contrary, if we naively expand the
BW factors in powers of the coupling constant, nonintegrable
singularities will be generated. This fact constitutes a well-known
problem of the determination of the cross-section in the case of
productions and decays of unstable particles. In the framework of
systematic expansion of amplitude in powers of the coupling
constant this problem, seemingly, is unsolvable.\footnote{Let us
remind that the CMS cannot be considered as a systematic approach
for the calculation of amplitude beyond the NLO \cite{CMS2}.}
However, at the level of the cross-section, a solution to the
problem does exist.

The basic idea is to consider the singularities in the
cross-section in the sense of distributions. In a mathematically correct
way, the problem may be stated as a problem of asymptotic expansion of
BW factors in powers of the coupling constant in the sense of
distributions. In the case of a separately taken BW factor with smooth
weight, a solution to the problem is well-known \cite{Tkachov}. Namely,
the expansion is beginning with the $\delta$-function, which
corresponds to the narrow-width approximation. The naive
Taylor-expansion terms are supplied with the principal-value
prescription for the poles. Strongly nontrivial contribution appear as
a sum of the delta-function and its derivatives with coefficients
$c_{n}$, which are polynomials in the coupling constant~$\alpha$. Within
the NNLO, the expansion looks as follows:
\begin{eqnarray}\label{not3}
\displaystyle \rho(s) & \equiv & \frac{M\Gamma_0}{\pi} \; {\Bigl|s
- M^2 + \Sigma(s)\Bigr|^{-2}}
\\[0.5\baselineskip]
\displaystyle & = & \delta(s\!-\!M^2) \,+\,\frac{M \Gamma_{0}}{\pi}
\, PV \! \left[\,\frac{1}{(s-M^2)^2} -\,
\frac{2\alpha\,\Re\Sigma_1(s)}{(s\!-\!M^2)^3}\,\right] \nonumber
\\
&& + \sum\limits_{n\,=\,0}^2 c_{n}\,
  \frac{\mbox{\small ($-$)}^{n}}{n!}\,\delta^{(n)}(s\!-\!M^2) +
  O(\alpha^3)\,.\nonumber
\end{eqnarray}
Here $M$ is the renormalized mass of the unstable particle,
$\Gamma_{0}$ is its Born width, $\Sigma = \alpha \,\Sigma_1 +
\alpha^2 \,\Sigma_2 + \alpha^3 \,\Sigma_3 + \cdots$ is the
self-energy in the scalarized propagator. Coefficients $c_n$
within the NNLO are determined by three-loop self-energy contributions
and by their derivatives, all taken on-shell. The structure of the
contributions is such that in the OMS-type schemes of the UV
renormalization the real self-energy contributions appear without
the derivatives or with the first-order derivative only. Such
contributions are determined by the renormalization conditions. In
the unstable-particles case a convenient scheme of the UV
renormalization is the $\overline{\mbox{OMS}}$ or
``pole'' scheme \cite{OMS-bar,Sirlin}, defined by equating the
renormalized
mass of unstable particle to real component of the pole of its
propagator. This implies that the $\overline{\mbox{OMS}}$ mass
coincides with the observable mass, which is independent of the UV
renormalization scheme (and which is gauge-invariant). The second
renormalization condition is defined by equating imaginary component of
the on-shell self-energy to imaginary component of the pole.
Coefficients $c_n$ in this scheme are determined as follows \cite{MPT}:
\begin{equation}\label{not4}
c_0  =  - \, \alpha \, \frac{I_2}{I_1} + \alpha^2
\left[\frac{I_2^2}{I^2_1} - \frac{I_3}{I_1} - (I_{1}^{\,\prime})^2
\right],\qquad c_1 = 0, \qquad c_2 = -\, \alpha^2 I^2_1 \,,
\end{equation}
where $I_{k} = \Im\,\Sigma_{k}(M^2)$, and $I_{1}^{\,\prime} =
\Im\,\Sigma^{\,\prime}_{1}(M^2)$.

The next ingredient of the MPT is the analytic regularization of the
kinematic factor via the substitution $[\lambda (s,s_{1},s_{2})]^{1/2}
\to [\lambda (s,s_{1},s_{2})]^{\nu}$ \cite{MPT}. The regularization is
necessary to impart enough smoothness to the weight at the BW factors in
formula (\ref{not2}). Under the regularization, by representing the
remaining contribution to the weight in the form of Taylor expansion
with a remainder, we can do analytic calculation of singular integrals
irrespective of details of the definition of the weight.
Further, by putting $\nu = 1/2$, we obtain finite expressions with
the expansion being asymptotic \cite{MPT}. Thereby the problem
is reduced to numerical calculations only.

Now we turn to the definition of the model for carrying out the
calculations. Acrually, we define the model in many ways by analogy
with \cite{Ntt}. First of all, we note that the MPT-expansion of
the BW factor based on the full propagator of $W$-bosons coincides
within the NNLO with that based on the following modelling propagator:
\begin{eqnarray}\label{not5}
\Delta^{-1}(s) & = & s \, - \, M^2 \, + \, \alpha \, \Re
\Sigma_1(s) \, + \, \mbox{i}\,\alpha \, \Im \Sigma_1(s)
\nonumber\\[0.5\baselineskip]
&& + \; \alpha^2 \left[R_2 + \mbox{i}\, I_2 + (s-M^2) (R'_2 +
\mbox{i}\, I'_2)\, \right] \: + \: \alpha^3 \left(R_3 + \mbox{i}\,
I_3 \right).
\end{eqnarray}
Here $R_{k} = \Re\,\Sigma_{k}(M^2)$ and $R_{k}^{\,\prime} =
\Re\,\Sigma^{\,\prime}_{1}(M^2)$. When considering in the
sense of conventional functions, a discrepancy between propagator
(\ref{not5}) and the full propagator turns out to be beyond the NNLO,
as well. Below we consider propagator (\ref{not5}) as the full or
``exact'' propagator in the model. The $\Re \Sigma_1(s)$ and $\Im
\Sigma_1(s)$, we define by means of direct calculations in the SM.
However, for sake of simplicity we restrict ourselves by
consideration of only quark and lepton contributions. In this way we
avoid the IR divergences generally arising at calculating $\Re
\Sigma_{1}(s)$. For the definition of the on-shell self-energy
contributions in (\ref{not5}), we take advantage of the UV
renormalization conditions
(we use the $\overline{\mbox{OMS}}$ scheme) and the conditions of
unitarity. In this way we obtain (see details in \cite{Ntt} and
\cite{OMS-bar})
\begin{eqnarray}\label{not6}
& R_2 = -I_1 I'_1 \,,
\qquad R'_2= - I_1 I''_1 / 2 \,, \qquad
 R_3 = - I_2 I'_1 - I_1 I'_2 + I^2_1 R''_1 / 2\,,&
\\[0.5\baselineskip]\label{not7}
&\alpha I_1 = M\Gamma_0\,, \qquad
 \alpha^2 I_2 = M \Gamma_1\,, \qquad
 \alpha^3 I_3 = M \Gamma_2 + \Gamma_0^3/(8M)\,,&
\end{eqnarray}
where $\Gamma_0$, $\Gamma_1$, $\Gamma_2$ are the Born-, one-loop, and
two-loop contributions to the width. The $I'_2$, we determine with the
aid of an approximate relation $\alpha^2 I'_2 = (\Gamma_1/\Gamma_0)\,
\alpha I'_1$.

Further we proceed to the definition of the amplitude squared $\Phi$ in
formula (\ref{not2}). For the purpose of testing only of the convergence
properties of the MPT expansion, we determine $\Phi$ in the Born
approximation. In the main we follow \cite{Gentle}, except
$Z$-boson propagator. Recall that in the Born approximation the
$Z$-boson propagator arises immediately owing to $e^{+} e^{-}$
annihilation, and that it makes one-particle reducible contribution.
Therefore this contribution becomes resonant if $s'$ in formula
(\ref{not1}) falls on $s' = M_Z^2$. In the latter case the mentioned
propagator of $Z$-boson in $\Phi$ is to be MPT expanded.\footnote{Let
us remark that the corrections to $\Phi$ outside the above mentioned
$Z$-boson propagator, constitute one-particle irreducible contributions.
Therefore they are non-resonant and should be considered by means of
conventional perturbation theory.}
Unfortunately, the loop corrections to $Z$-boson propagator
without simultaneous consideration of the radiative corrections to
vertices, will lead to violation of unitarity at high energies.
For this reason, we restrict the energy range in which we consider
corrections to $Z$-propagator. Namely, we determine $Z$-boson
propagator by formulae (\ref{not5})-(\ref{not7}) at $s<s_Z$, and
as a free propagator at $s > s_Z$, where $s_Z$ is a parameter such that
$s_Z > M_Z^2$. By means of this trick we, in a proper way, take into
consideration the resonant contributions of $Z$-boson
propagator, and simultaneously preserve the gauge cancellations
and the unitarity at high energies. (Let us remember that this is
a model consideration. In the approach with the taking into
consideration of all radiative corrections, both in the vertices
and in the $Z$-boson propagator, the gauge cancellations and,
consequently, the unitarity should automatically be maintained.
This should be so, since we construct a complete asymptotic
expansion in powers of the coupling constant.)

For completing the definition of the model, we must determine also
the soft-photon factor $(1+\delta_{c})$ in formula (\ref{not2}).
Let us remember that we have omitted photon contributions in the
self-energies of $W$-bosons. So we have to ignore those
soft-photon contributions whose IR-divergent contributions are to be
cancelled in the cross-section. For this reason, we consider only
Coulomb soft-photon contributions in $\delta_{c}$. Concretely, we
take advantage of the well-known formula for the Coulomb factor in the
one-photon approximation, which includes appropriate off-shell
and finite-width effects, see details in \cite{MPT}.

Now the model is completely determined, and the cross-section in the
model may be straightforwardly calculated. We call the outcome the
``exact result in the model''. After the MPT-expansion, the
cross-section is represented in the form $\sigma(s) =
\sigma_{LO}(s) \,+\, \alpha\,\sigma_{1}(s) \,+\, \alpha^2
\sigma_{2}(s)$, where $\sigma_{LO}$ is the cross-section in the LO
approximation, $\alpha\,\sigma_{1}$ and $\alpha^2 \sigma_{2}$ are
the NLO and NNLO corrections, respectively. So, the $\sigma_{NLO} =
\sigma_{LO} + \alpha\,\sigma_{1}$ and~$\sigma_{NNLO} = \sigma_{LO} +
\alpha\,\sigma_{1} + \alpha^2 \sigma_{2}$ determine the NLO and the NNLO
approximations.

\section{Numerical results}\label{num}

For numerical calculations, we use parameters: $M_W = 80.40 \:
\mbox{GeV}$, $M_Z = 91.19 \: \mbox{GeV}$, $m_t = 175 \: \mbox{GeV}$,
and the other quarks and leptons are massless. To parametrize the
amplitude squared $\Phi$, we use effective coupling determined as
\begin{equation}\label{num1}
\alpha_{\mbox{\tiny \it G}_{\mu}} = \sqrt{2} \, G_{\mu} M_W^2
s_W^2/\pi,
\end{equation}
where $s_W^2 = 1-M_W^2/M_Z^2$, $G_{\mu}$ is the Fermi constant, $G_{\mu}
= 1.16637 \times 10^{-5}$~GeV$^{-2}$. For the calculation of relative
corrections, we use $\alpha = 1/137.036$. The parametrization of this
kind was used e.g.~in \cite{CMS2}. Since we consider the widths
as the relative corrections, the mentioned parametrization implies
large electroweak (EW) corrections to the widths. To estimate
specifically the $O(\alpha)$ EW corrections, we use results of
\cite{Denner} and \cite{Zwidth}. The $O(\alpha_s)$
corrections, we obtain in the ``naive'' approach, extracting them from
the corresponding factor $(1+\alpha_s/\pi)$ with $\alpha_s = 0.12$. In
view of the lack of complete results on the two-loop corrections to the
widths, we estimate they as the differences between the current values
$\Gamma^W = 2.082$~GeV, $\Gamma^Z = 2.534$~GeV \cite{PDG} and the sums
of the appropriate Born and one-loop results. In this way we obtain
\begin{eqnarray}\label{num2}
 \Gamma^W_{0}  &=& 1.977 \; \mbox{GeV} \,, \qquad\;\:\:\,
 \Gamma^Z_{0} \;=\; 2.362 \; \mbox{GeV} \,, \nonumber
\\[0.3\baselineskip]
 \Gamma^W_{1} &=& 0.102 \; \mbox{GeV} \,, \qquad\;\;\,\,
 \Gamma^Z_{1} \;=\; 0.169 \; \mbox{GeV} \,,
\\[0.3\baselineskip]
 \Gamma^W_{2} &=& 0.006 \; \mbox{GeV} \,, \qquad\;\;\,\,
 \Gamma^Z_{2} \;=\; - 0.036 \; \mbox{GeV} \,. \nonumber
\end{eqnarray}
Another way to estimate the two-loop corrections is to consider
they as equal to the two-loop QCD corrections extracted from the
factor $(1+\alpha_s/\pi+1.409 \,\alpha_s^2/\pi^2)$ \cite{PDG}. In
this way we obtain $\Gamma^W_{2} = \Gamma^Z_{2} = 0.003$~GeV. Below we
use the latter option for checking stability of the outcomes, but as the
main estimate we consider (\ref{num2}). Recall that we use the widths
for the determination of the ``exact'' propagator (\ref{not5}) and of
the coefficients $c_n$ in the MPT expansion of the BW factors. Parameter
$s_Z$ in the definition of $Z$-boson propagator, we determine on the
basis of the condition of the most smooth ``sewing'' solutions above and
below $s_Z$.
This gives $\sqrt{s_Z} = 130$--150~GeV, and the ultimate result is
practically independent from the concrete choice. Finally, we put
$\sqrt{s_{i\,\mbox{\scriptsize min}}} =30$~GeV in formula (\ref{not2}).
The calculations, we carry out on the basis of FORTRAN code with double
precision written in accordance with results of \cite{MPT}.

\begin{figure}[tp]
\hbox{ %\hspace*{60pt}
       \epsfxsize=420pt \epsfbox{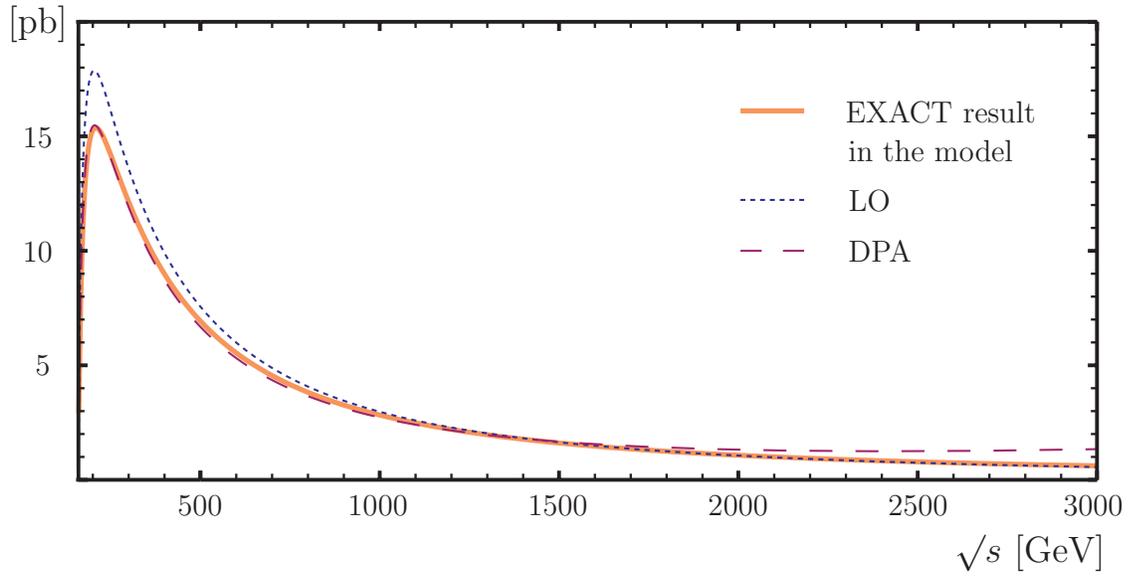}}
%\vspace*{-0.5\baselineskip}
\caption{\small Total cross-section: the exact result in the model, the
results in the LO in the MPT (the narrow-width approximation), and in
the DPA.\label{Fig1}}
\vspace*{-\baselineskip}
\end{figure}

\begin{figure}[tp]
\hbox{ %\hspace*{60pt}
       \epsfxsize=420pt \epsfbox{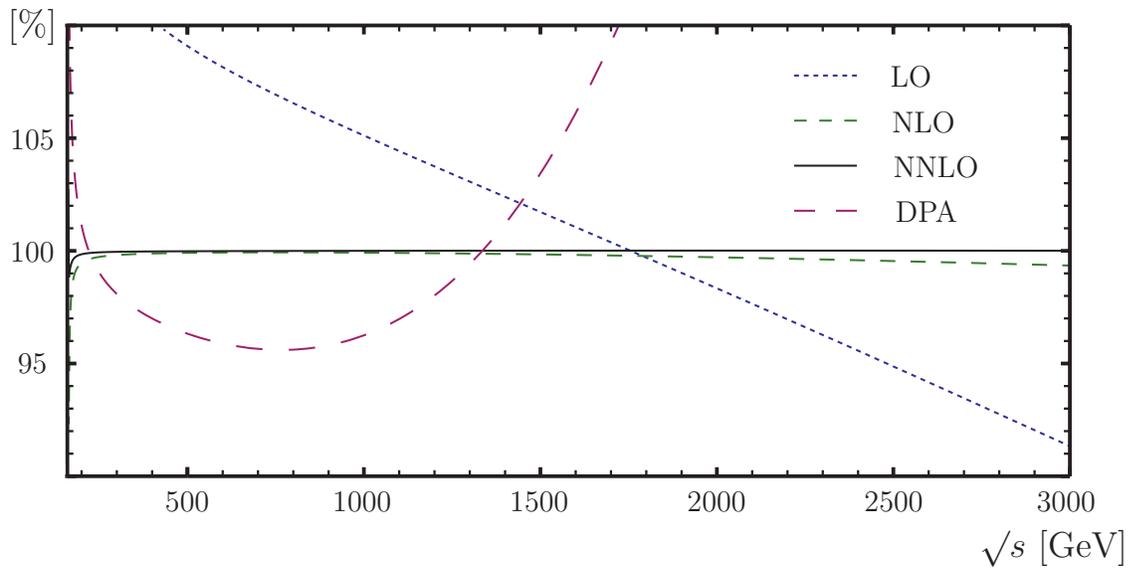}}
%\vspace*{-0.5\baselineskip}
\caption{\small Percentages with respect to ``exact'' cross-section
calculated in the LO, NLO, and in the NNLO approximations in the MPT,
and in the DPA.\label{Fig2}}
\vspace*{-\baselineskip}
\end{figure}

\begin{figure}[p]
\hbox{ %\hspace*{60pt}
       \epsfxsize=420pt \epsfbox{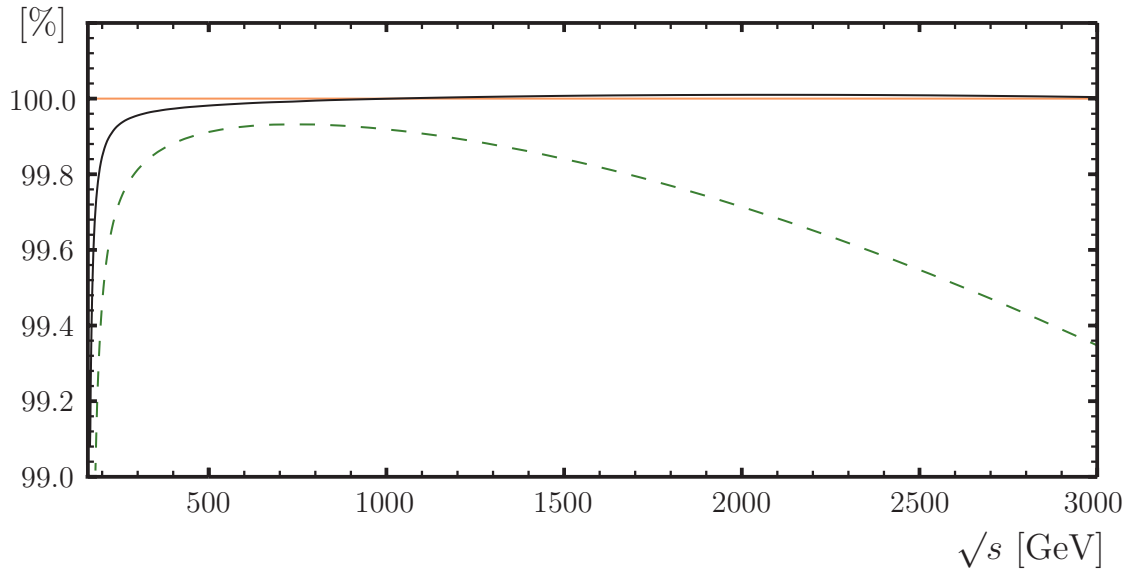}}
\vspace*{-0.5\baselineskip}
\caption{\small Percentages to the exact cross-section, calculated in
the NLO and NNLO approximations in the MPT. The horizontal line at
100.0\% means the exact result. The notation for the curves is the same
that in Fig.~\ref{Fig2}.\label{Fig3}}
\vspace*{-\baselineskip}
\end{figure}

\begin{figure}[h]
\hbox{ %\hspace*{60pt}
       \epsfxsize=420pt \epsfbox{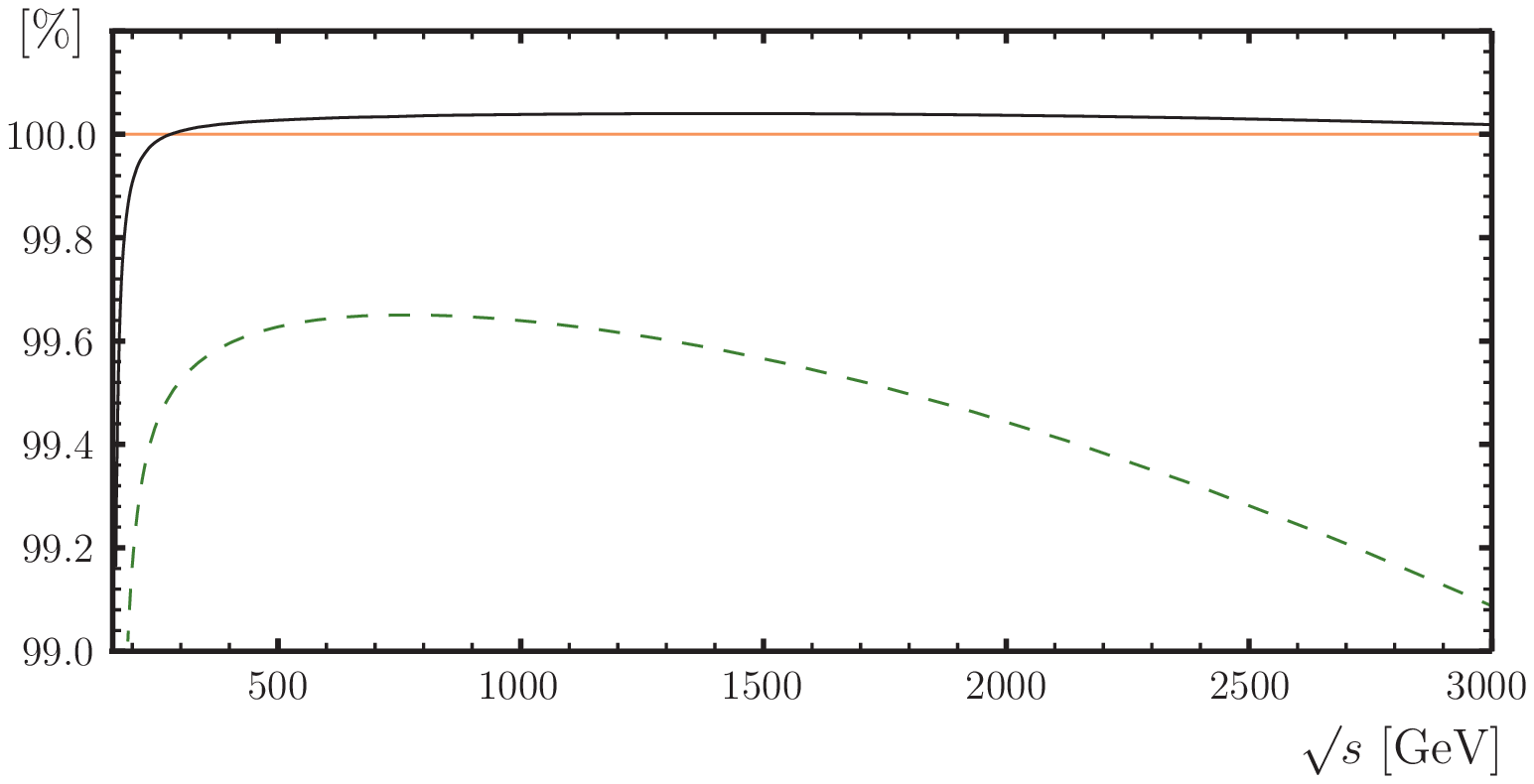}}
\caption{\small The same as in Fig.~\ref{Fig3} but with the other input
for widths (see explanation in the text).\label{Fig4}}
\end{figure}

The outcomes of the calculations are presented in
Figs.~\ref{Fig1}--\ref{Fig3} and in Table~1. In Fig.~\ref{Fig1},
the thick curve shows the exact result for the total cross-section, the
dotted and long-dashed curves show the cross-section in the LO
approximation of the MPT and in the DPA, respectively. Recall that
the LO in the MPT coincides with the narrow-width approximation. The DPA
is constructed by substitution of
$\Phi(s;M^2,M^2)(1+\delta_{c}(s;M^2,M^2))$ for
$\Phi(s;s_1,s_2)(1+\delta_{c}(s;s_1,s_2))$ and $s - M^2 +
\mbox{i}\,\Gamma$
for $\Delta^{-1}(s)$ in the expression for the exact result. The
curves for the NLO and NNLO are not presented in Fig.~\ref{Fig1}, since
they merge with the curve for the exact result at a given scale.
Instead, they are shown in Figs.~\ref{Fig2} and \ref{Fig3}, where
the percentages with respect to the exact result are presented. In
Fig.~\ref{Fig3} the results of Fig.~\ref{Fig2} are repeated with greater
scale on vertical axis. In Table~1 the outcomes are presented in
numerical form at the characteristic energies at the ILC. In the last
column the numbers in parenthesis represent uncertainties in the last
digits arising due to computations. In other columns the uncertainties
are omitted as they are beyond the precision of the presentation of
data. The estimation of uncertainties is made in accordance with the
algorithm described in \cite{ACAT}. The results of checking
stability of the outcomes with respect to the variation of the
corrections to the widths (see explanation in the previous paragraph)
are shown in Fig.~\ref{Fig4}.


\begin{table}[p]
\begin{center}
\begin{tabular}{ c  c c c c  }
\hline
\hline\noalign{\medskip} \\[-6mm]
 $\quad \sqrt{s}$ (TeV) $\quad\;\;$
 & $\quad\; \sigma_{\,\mbox{\tiny EXACT}} \quad$      &
 $\qquad \sigma_{LO} \qquad$        & $\qquad \sigma_{NLO} \qquad$ &
 $\quad \sigma_{NNLO}   \qquad$ \\
\hline\noalign{\medskip} \\[-6mm]
 0.2                               & 15.258         &
 17.839          &  15.175         & 15.235(2)          \\%[1mm]
                                   & {\small 100\%} &
{\small 116.92\%}& {\small 99.46\%}& {\small 99.85(1)\%} \\[2mm]
 0.5                               & 6.9355         &
 7.5657          &  6.9294         & 6.9342(7)          \\%[1mm]
                                   & {\small 100\%} &
{\small 109.09\%}& {\small 99.91\%}& {\small 99.98(1)\%}\\[2mm]
 1                                 & 2.8286         &
 2.9733          &  2.8263         & 2.8285(3)          \\%[1mm]
                                   & {\small 100\%} &
{\small 105.12\%}& {\small 99.92\%}& {\small 100.00(1)\%} \\[2mm]
 3                                 & 0.61023        &
 0.55733         &  0.60625        & 0.61026(6)         \\%[1mm]
                                   & {\small 100\%} &
{\small 91.33\%} & {\small 99.35\%}& {\small 100.00(1)\%} \\[2mm]
 5                                 & 0.391580         &
 0.24353         &  0.31047        & 0.31566(3)         \\%[1mm]
                                   & {\small 100\%} &
 {\small 77.11\%} & {\small 98.31\%} & {\small 99.95(1)\%} \\
\noalign{\smallskip}\hline\hline
\end{tabular}
\caption{\small The results of the calculation of the total
cross-section in pb and in \% with respect to the exact result in the
model.}
\end{center}\vspace*{-0.9\baselineskip}
\end{table}

The presented outcomes exhibit stable behavior of the NLO approximation
and very stable behavior of the NNLO approximation in the energy region
beginning with approximately 200~GeV. At the same time, the accuracy of
the NNLO approximation in this energy region (more precisely, above
220~GeV) is less than one per-mille. These results remain valid
at varying the input-data for the widths $\Gamma^W_{2}$ and
$\Gamma^Z_{2}$, cf.~Fig.~\ref{Fig3} and Fig.~\ref{Fig4}. 
The corresponding shift of the NNLO approximation with the changing of
the widths constitutes a few of $10^{-4}$ in relative units. The
concurrent shift of the NLO curve exactly follows the shift of the
``exact'' solution determined by propagator (\ref{not5}), and in
relative units constitutes a few of $10^{-3}$. As for the DPA, it has
rather non-stable behavior and at high energies it becomes openly bad.
In particular, at the energies above 1.5~TeV the DPA becomes much worse
even of the narrow width approximation (the LO approximation in the
MPT).

In the end of this section it is worth noticing that at the approaching
to the threshold of $W$-pair production, the accuracy and stability of
MPT rapidly become worse. Actually this behavior was predicted in
\cite{MPT} and was observed at previous
calculations \cite{N-tt,Ntt}. To prevent this difficulty another mode of
the MPT near threshold should be applied, which includes Taylor
expansion of $\sigma(s)$ not only in powers of $\alpha$, but also in
powers of the distance between $s$ and the threshold \cite{MPT}. A study
of the behavior of the MPT in this mode is a task for future
investigations.

\section{Discussion and conclusions}\label{Summary}

We have tested applicability of the MPT for description of $W$-pair
production in $e^{+} e^{-}$ annihilation. We have found that MPT
provides stability of outcomes at the energies beginning with
approximately 200~GeV. By means of model calculations we have shown that
at the mentioned energies the NNLO approximation in the MPT provides
better than one per-mille precision of the description of the total
cross-section.

The obtained precision exceeds that observed in the case of the
top-quark pair production \cite{Ntt}. The difference is
caused by the the lower values of the corrections to the width of
$W$-boson as compared to the corrections to the width of the
top-quark (and to a lesser degree by the opposite sign of the
corrections). Specifically, if we use the corrections to the width of
the top-quark in place of those for the $W$-boson (in relative units) in
the present calculations, the outcomes for the cross-section
in relative units are obtained almost identical to those in the
case thoroughly with the top-quark pair production. On the one hand,
this confirms the applicability of the MPT in the presence of the large
gauge cancellations that occur in the case of the $W$-pair production.
On the other hand, this confirms the previous observation that the
results of the MPT calculations in relative units depend rather weakly
on the particular form of the test function (the amplitude squared
without the resonant factors), but depend mainly on the structure of the
resonant factors \cite{Ntt}. In total, the above observations mean that
the modification of the test function due to turning on the loop
corrections, in relative units should not lead to any significant
modifications in the cross-section. So, our main results should remain
in force in the presence of the loop corrections, in particular, the
result about the better than one per-mille precision of the description
of the cross-section.

In summary, we conclude that the MPT is really a good candidate for the
description of $W$-pair production at the ILC. For a definitive
conclusion, of course, a more comprehensive analysis must be made, in
particular the analysis of the angular distributions. A study of
complete realistic processes, such as CC10, CC11 etc., in the framework
of an appropriate model must be made, as well. The mentioned
investigations will be carried out elsewhere.

\end{document}